\documentstyle[preprint,eqsecnum,aps,epsfig]{revtex}

\renewcommand{\baselinestretch}{1.0}
\begin{document}

\title{\vspace{1.5cm} Classification of higher order diagrams in
non-equilibrium theory, and the removal of pinch singularities}
\author{D. S. Isert\footnote{E-mail: D.Isert@tphys.uni-heidelberg.de} 
and S. P. Klevansky\footnote{Current address: DB AG,
Kleyerstr.27, D-60236 Frankfurt a.~M., Germany}}
\address{ Institut f\"ur Theoretische Physik,\\
Philosophenweg 19, D-69120 Heidelberg, Germany}

\maketitle
\vspace{1cm}

\renewcommand{\baselinestretch}{1.3}
\footnotesize\normalsize
\begin{abstract}
The non-equilibrium two loop self-energy is reexamined in the framework of a
scalar quark and gluon model, with specific attention to terms which do not
give rise to the standard two $\to$ two particle collisional terms in the
semi-classical Boltzmann equation. It is shown that most of these terms
contribute to renormalization of component lines at a lower level, rendering
the theory correct to $g^4m^4$. This result can be generalized to a higher
number of loops. The remaining terms, which do not fall into any physical
category, are shown explicitly to vanish. We then examine the possibility
that pinch singularities could arise in this theory, and demonstrate that
this is not so for the case of equilibrium and small deviations from
equilibrium, at the two loop level. 
\end{abstract}

\clearpage

\section{Introduction}
\label{sec:intro}

Due to the short time span of a heavy-ion collision, it is believed that
a non-equilibrium description of such a collision may be necessary to
give a complete and adequate description of the particle evolution.   
Several reviews of the non-equilibrium approach in general may be found in
\cite{mal}. For heavy ion collisions,
heuristic models have been developed, which are guided
by the approach of Boltzmann, that presumes that only two body collisions
can occur.   These can then be generalized in an heuristic fashion
\cite{mueller}. On
the other hand, from the field theoretical point of view, there have been
many attempts to generate the correct and appropriate non-equilibrium theory 
for a given Lagrangian (see, for example \cite{geiger}.   
Often these stop at the mean-field level, due to
the inherent complexity of the non-equilibrium formalism per se.   In some
previous works, the two-body collisional term has been calculated for a 
static potential interaction, and under many simplifying assumptions in
\cite{ogu}.
Although a graphical series was given for the two loop self-energy following
from a Walecka type interaction in \cite{heinz}, the physical interpretation 
of all the terms that occur in this series or direct expressions for them was 
not given. Note that the difficulty lies in the fact that the
non-equilibrium formalism generates far more diagrams of a given topology
than does the equilibrium zero temperature or Matsubara finite temperature
formalism which generates Feynman graphs in the standard way. 
In a broad sense, the ultimate goal which we have, is to obtain a complete 
physical understanding of {\it all} the additional terms (over and above
those which are expected from a Feynman diagram point of view)
that are generated in the non-equilibrium formalism, in
order that a precise formulation in, for example, the study of
non-equilibrium evolution of distribution functions in QCD, rather than
simply a heuristic approach, can be made. At this stage, however, we
restrict ourselves to far simpler systems.

In this paper, we examine the collisional integral for a scalar quark-gluon
model in the Hartree-Fock approximation and for the two loop diagrams in 
greater detail than before \cite{us}, placing specific emphasis on the extra 
terms that are naturally generated in the non-equilibrium formalism. 
In this study, (a) we are able to show that two-loop diagrams that were 
previously ignored \cite{us} as they do not contribute directly to the 
two-particle scattering terms expected from classical Boltzmann theory, 
do in fact have an important physical relevance:  
all of these remaining graphs are required to correct the {\it lower} order 
tree level graphs which give rise to absorption or radiation processes and
which also occur as matrix elements in the Boltzmann equation to one loop, 
implying that a simulation of a non-equilibrium process with this or similar 
models requires inclusion of such terms.
(b) Our study is carried out within the Schwinger-Keldysh formalism
\cite{schwing} for non-equilibrium processes. Within this (and other) 
formulations of real-time finite temperature processes \cite{umezawa1}, 
the issue of pinch singularities arises (see for example 
\cite{altherr,greiner,dadic}). 
This addresses the question as to whether the theory is well-defined or not, 
in that it possesses divergences that arise apparently from integration along 
the real axis of products of retarded and advanced Green functions $G_RG_A$. 
Such products occur naturally in the socalled 'retarded-advanced' (RA) 
version of the non-equilibrium 
formalism, which is an alternative way of presenting the Schwinger-Keldysh
Green functions \cite{landau}. Alternately, pinch singularities manifest 
themselves as products of delta functions within the standard representation
which we follow here. Within our model, we investigate the occurrence
of pinch singularities in the series for the self-energy up to two loops   
and come to the conclusion that no pinch singularities occur.

To be specific, we utilize the scalar partonic model \cite{polk,forsh}, 
defined by the Lagrangian
\begin{eqnarray}
{\cal L} &=& \partial^\mu \phi^\dagger{}^{i,l}\partial_\mu\phi_{i,l} +
\frac 12 \partial^\mu\chi_{a,r}\partial_\mu\chi^{a,r} - \frac{m^2}2
\chi_{a,r}\chi^{a,r} \nonumber\\
&& - gm\phi^\dagger{}^{i,l}(T^a)^j_i(T^r)^m_l\phi_{j,m}
\chi_{a,r} - \frac{gm}{3!} f_{abc}f_{rst} \chi^{a,r}\chi^{b,s}\chi^{c,t}
\label{e:lagr}
\end{eqnarray}
in which scalar quark fields $\phi^{i,l} $ are coupled to scalar gluonic 
fields $\chi^{a,r}$ via a cubic interaction. The color indices $i,l=1..N_c$, 
while $a,r=1..(N_c-1)$, reflects the product groups 
$SU(N_c)\times SU(N_c)$.

\section{Collision Term to two loops}
\label{sec:coll}
In order to recover the semi-classical extension of the Boltzmann equation,
it is necessary to evaluate constructs of the form \cite{ogu,us}
\begin{eqnarray}
I_{\rm coll} &=&  \Pi^{-+}(X,p)D^{+-}(X,p)- \Pi^{+-}(X,p) D^{-+}(X,p)
\nonumber \\
&=& I_{\rm coll}^{\rm gain} - I^{\rm loss}_{\rm coll}, 
\label{e:ifunction}
\end{eqnarray}
where $D^{ij}(X,p)$, $i=+,-$ are Green functions of the standard 
Schwinger-Keldysh kind, and which are given in Appendix A. The transport and
constraint equations are briefly summarized in this Appendix also.
The $\Pi^{ij}$ are the Wigner transformed self-energies of the associated
particle.   To be specific, we consider for the moment the self-energies and
Green functions pertaining to the quark sector, $\Pi^{ij}(X,p) =
\Sigma^{ij}(X,p)$ and $D^{ij}(X,p) = S^{ij}(X,p)$. The analysis for the
gluonic sector follows then simply by inspection. Let us  
consider first the loss term. This means we need to calculate 
$\Sigma^{+-}(X,p)$. The quark self-energy to one loop is shown in Fig.1a)
which is the Fock term, and to two loops in Fig.2. As there are many graphs
at the two loop level, we make a structural classification of these
according to their topology. We identify rainbow graphs (R), ladder graphs 
(L), cloud graphs (C), exchange graphs (E) and quark-loop graphs (QL). 
In \cite{us}, we were able to show explicitly that the Fock term of Fig.1a) 
leads to the requirement that the process $q\bar q \to g$, as given by 
Fig.1b), be incorporated into the associated Boltzmann equation at this
level. We have also shown that the seven diagrams R a),
L a), C a), C b), E a), E b) and QL a) are all essential for
constructing the direct cross-sections of the processes $qq\to qq$, 
$q\bar q\to q\bar q$, $q\bar q \to gg$, and $qg \to qg$ which are shown in
Fig3, and which then appear in the newly calculated Boltzmann equation that
should be correct to the two loop level. Two questions now arise naturally.
Firstly, one may ask what the purpose of the remaining thirteen graphs is,
and secondly, more subtly, would they harbour divergences such as arise
through pinch singularities, thereby rendering the theory ill-defined. In a 
previous paper, the remaining thirteen two-loop self-energy diagrams shown 
in Fig.2 
were argued heuristically to be 'non-leading', and graphs of this type have 
up to now simply been ignored. Here, we wish to demonstrate precisely that 
these graphs in fact either vanish or contribute to corrections of order 
$g^3m^3$ to the lower order process $q\bar q \to g$. 
To demonstrate this, we arbitrarily examine the set of quark-loop diagrams.
The QL a) graph of Fig.2 leads directly to the $qq$ and $q\bar q$ cross
sections of Fig.3. On the other hand, the 
three graphs, QL b)-d) of Fig.2 which were {\it not} required for
the evaluation of the cross-sections as derived for Fig.3.
Let us examine each of these terms in turn.
We commence with the diagram QL b) of Fig.2 which is given by
\begin{eqnarray}
\Sigma^{(2)+-}_{{\rm q-Loop,b)}} (X,p)&=&-g^4m^4F_{{\rm q-Loop}}^2           
\int\!\frac{d^4\!p_1}{(2\pi)^4}
\frac{d^4\!p_2}{(2\pi)^4}\frac{d^4\!p_3}{(2\pi)^4}\frac{d^4\!p_4}{(2\pi)^4}
(2\pi)^4\delta^{(4)}(p-p_1-p_2)\nonumber\\
&&\times (2\pi)^4
\delta^{(4)}(p_2-p_3+p_4)\,S^{+-}(X,p_1)\,G^{+-}(X,p_2)\nonumber\\
&&\times G^{++}(X,p_2)\,S^{++}(X,p_3)\,S^{++}(X,p_4),
\label{e:qloopb}
\end{eqnarray}
where $F_{\rm q-Loop}^2$ is the color factor for the two color $SU(N_c)$
groups.
The corresponding loss term of the collision integral to this self-energy is 
defined by
\begin{equation}
J_{\rm coll,q-Loop,b)}^{(2) \rm loss}=-i\frac{\pi}{E_p}\Sigma^{(2)+-}_{{\rm
q-Loop,b)}}(X,p^0=E_p,\vec p)f_q(X,\vec p).
\label{e:j}
\end{equation}
In this expression, the product $iS^{+-}(X,p_1)
iG^{+-}(X,p_2)f_q(X,\vec p)$ occurs. Inserting the quasiparticle
approximation for the Green functions that are given in appendix A,
we obtain for this product the sum of four terms:
\begin{eqnarray}
T_1&=& \frac{\pi}{E_1}\frac{\pi}{E_2}\delta(E_1-p_1^0)
\delta(E_2-p_2^0)\bar f_q(X,p_1)\bar f_g(X,p_2) f_q(X,\vec p)\nonumber\\
T_2&=& \frac{\pi}{E_1}\frac{\pi}{E_2}\delta(E_1-p_1^0)
\delta(E_2+p_2^0)\bar f_q(X,p_1)f_g(X,-p_2) f_q(X,\vec p)\nonumber\\
T_3&=& \frac{\pi}{E_1}\frac{\pi}{E_2}\delta(E_1+p_1^0)
\delta(E_2-p_2^0)f_{\bar q}(X,-p_1)\bar f_g(X,p_2) f_q(X,\vec p)\nonumber\\
T_4&=& \frac{\pi}{E_1}\frac{\pi}{E_2}\delta(E_1+p_1^0)
\delta(E_2+p_2^0)f_{\bar q}(X,-p_1)f_g(X,-p_2) f_q(X,\vec p).
\end{eqnarray} 
By attributing $f$ to incoming particles and $\bar f$ to outgoing ones, we
see that $T_1...T_4$ correspond to the processes $q\to qg$, $qg\to q$,
$q\bar q\to g$ and $q\bar q g\to${\O}. Since the quarks are massless while
the gluons are endowed with a finite mass, the processes corresponding to 
$T_1$, $T_2$ and $T_4$ are kinematically forbidden as also occurred
in the discussion of the Fock term in \cite{us}.
One thus has one remaining non-vanishing contribution 
$iS^{+-}(X,p_1)iG^{+-}(X,p_2)f_q(X,\vec p)=T_3$.
This product is now inserted into Eq.(\ref{e:j}). In the resulting
expression, we interchange $p_1$ with $-p_1$ and on performing the 
$p_1^0$, $p_2^0$ and $p_4$ integrations, we find the result 
\begin{eqnarray}
J_{\rm coll,q-Loop,b)}^{(2) \rm loss}&=&-i\frac{\pi}{E_p}g^4m^4F_{{\rm
q-Loop}}^2 \int\!\frac{d^3\!p_1}{(2\pi)^3 2E_1}\frac{d^3\!p_2}{(2\pi)^3
2E_2}\frac{d^4\!p_3}{(2\pi)^4}(2\pi)^4\delta^{(4)}(p+p_1-p_2)\nonumber\\
&&\times G^{++}(X,p_2)\,S^{++}(X,p_3)\,S^{++}(X,p_3-p_2)
f_{\bar q}(X,\vec p_1) \bar f_g(X,\vec p_2) f_q(X,\vec p)\nonumber\\
\label{lossQLb1}
\end{eqnarray}
as the remaining contribution of the QL b) graph to the collision integral. 

Now, in Fig.4, we independently examine 
the Feynman diagrams for the process $q\bar q \to g$ up to order $g^3m^3$.
The scattering amplitude associated with Fig.4a) has purely a point-like
structure with color groups occurring:  
\begin{equation}
-i{\cal M}_{q\bar q\to g}^{a)}= -igm t_{ij}^a\otimes t_{lm}^r
\label{e:4a}
\end{equation}
while from Fig.4b), one has
\begin{eqnarray}
-i{\cal M}_{q\bar q\to g}^{b)}&=& g^3m^3 [t_{ji}^b{\rm tr}(t^bt^a)] \otimes
[t_{ml}^s{\rm tr}(t^st^r)]   G^{--}(X,p_2)\nonumber\\
&&\times\int\!\frac{d^4\!p_3}{(2\pi)^4}S^{--}(X,p_3)\,S^{--}(X,p_3-p_2),
\label{e:4b}
\end{eqnarray}
where $t^a_{ij}$ is the matrix of the color group in the representation of
the quarks. Now note that  
using the fact that $[iG^{--}]^{\dagger}= iG^{++}$ and
$F_{\rm q-Loop}=t_{ij}^at_{ji}^b{\rm tr}(t^bt^a)$, one can rewrite 
Eq.(\ref{lossQLb1}) in terms of these matrix elements, i.e.
\begin{eqnarray}
J_{\rm coll,q-Loop,b)}^{(2) \rm
loss}&=&\frac{\pi}{E_p}\int\!\frac{d^3\!p_1}{(2\pi)^3 2E_1} 
\frac{d^3\!p_2}{(2\pi)^32E_2} (2\pi)^4\delta^{(4)}(p+p_1-p_2)\nonumber\\
&&\times{\cal M}_{q\bar q\to g}^{a)}
[{\cal M}_{q\bar q\to g}^{b)}]^{\dagger}f_q(X,\vec p)f_{\bar q}(X,\vec p_1)
\bar f_g(X,\vec p_2),
\label{lossQLb2}
\end{eqnarray}
illustrating that the cross term between these two processes, denoted
symbolically as $a^{\dagger}b$, is derived from the self-energy diagram QL b)
of Fig.2.
The gain term can be obtained by replacing $f$ with $\bar f$ and
vice versa in Eq.(\ref{lossQLb2}).

In a similar fashion, the collision integral can be constructed from the 
quark-loop diagram QL c) in Fig.2. One obtains an expression for the loss term 
as in Eq.(\ref{lossQLb1}) with $iG^{++}iS^{++}iS^{++}$ replaced by the
combination $iG^{--}iS^{--}iS^{--}$. Again 
$J_{\rm coll,q-Loop,b)}^{(2)\rm loss}$ can be expressed by the scattering 
amplitudes of Eq.(\ref{e:4a}) and (\ref{e:4b}):
\begin{eqnarray}
J_{\rm coll,q-Loop,c)}^{(2) \rm
loss}&=&\frac{\pi}{E_p}\int\!\frac{d^3\!p_1}{(2\pi)^3 2E_1}
\frac{d^3\!p_2}{(2\pi)^32E_2} (2\pi)^4\delta^{(4)}(p+p_1-p_2)\nonumber\\
&&\times [{\cal M}_{q\bar q\to g}^{a)}]^{\dagger}
{\cal M}_{q\bar q\to g}^{b)} f_q(X,\vec p)f_{\bar q}(X,\vec p_1)
\bar f_g(X,\vec p_2),
\end{eqnarray}
i.e.~the second cross term $b^{\dagger}a$ required in building a
cross-section of the basic component a) and b) of Fig.4 is required.

In an analogous fashion, one can show that the rainbow diagrams R b) and c) 
lead to a collision integral containing ${\cal M}_{q\bar q\to g}^{a)}
[{\cal M}_{q\bar q\to g}^{f)}]^{\dagger}$ and the hermitian conjugate of this
product, the ladder diagrams Lb) and c) to a collision integral
containing ${\cal M}_{q\bar q\to g}^{a)}
[{\cal M}_{q\bar q\to g}^{c)}]^{\dagger}$ and its hermitian 
conjugate, the cloud diagrams C c) and d) to a collision integral containing 
${\cal M}_{q\bar q\to g}^{a)}[{\cal M}_{q\bar q\to
g}^{d)}]^{\dagger}$ and its hermitian conjugate, and finally the exchange
diagrams E c) and d) to a collision integral containing
${\cal M}_{q\bar q\to g}^{a)}[{\cal M}_{q\bar q\to
g}^{e)}]^{\dagger}$ and its hermitian conjugate.
Note that in this fashion, we are able to account for all mixed diagrams
that would occur in the construction of the $|{\cal M}_{q\bar q\to g}|^2$ up
to order $g^4m^4$, with the exception of the diagram g) of Fig.4.
This graph does not enter into the collision integral, as it 
is a renormalization diagram for the {\it incoming} quark, for which the
momentum $p$ is fixed externally. 

Returning to our explicit example of the quark-loop self-energy of Fig.2, one
sees that a simple graphical interpretation can be applied to each figure. A
rule in which all lines that are connected by $\pm$ and $\mp$ are cut in a
single path, separates the graphs QL a)-c) into their component matrix
elements. This is illustrated in Fig.5. This procedure, however, cannot be
applied uniquely to the graph QL d), nor for that matter to the remaining
graphs which are not required for construction of the mixed terms or direct
contributions to the cross-sections, i.e.~the graphs R d) and L d).
We are thus now left with the three graphs QL d), R d) and L d) which at
first sight fit into no apparent scheme, and which therefore may present
difficulties. 

We commence with the investigation of QL d). For this self-energy we obtain
an expression as in Eq.(\ref{e:qloopb}) with the product of the five Green 
functions replaced by 
$S^{+-}(X,p_1)[G^{+-}(X,p_2)]^2S^{-+}(X,p_3)S^{+-}(X,p_4)$.
In the quasi-particle approximation the off-diagonal Green functions
are on mass shell:
\begin{eqnarray}
p^2_1=p^2_3=p^2_4&=&0    \label{1}\\
p^2_2&=&m^2 \label{2}
\end{eqnarray}
In addition to this, the two $\delta$-functions for the energy-momentum
conservation of Eq.(\ref{e:qloopb}) have
to be fullfilled. Therefore we can write for example
\begin{equation}
0=p^2_3=(p_2+p_4)^2=2p_2p_4+m^2.
\end{equation}
Together with Eq.(\ref{2}) this leads to 
\begin{equation}
p_4=-\frac{1}{2}p_2\quad=>\quad p_4^2=\frac{1}{4}p_2^2=\frac{1}{4}m^2> 0 .
\end{equation}
This is inconsistent with Eq.(\ref{1}) and therefore 
$\Sigma^{(2)+-}_{{\rm q-Loop,d)}}$ has to vanish. 
A similar investigation of the graphs  R d) and L d) shows that they vanish 
for the same reason. 

So far we have derived the results for two loop self-energy diagrams in
detail. Let us summarize now the results for self-energy diagrams with more
than two loops. 
The result obtained in this section can be generalized easily to three and
more loops. Including the three loop self-energy into the collision term
leads to cross sections of all possible 
processes with two (three) partons in the initial state and three (two) 
partons in the final state. Furthermore, they give all additional corrections 
of order $g^6m^6$ to the process $q\bar q\to g$, e.g. the product of 
scattering amplitudes of the diagrams b)-f) in Fig.4, and also corrections 
to all scattering processes of two partons into two partons, $qq\to qq$, 
$q\bar q\to q\bar q$, $q\bar q\to gg$ and $qg\to qg$.

\section{Pinch singularities}
\label{sec:pinch}
To investigate the issue of pinch singularities, let us make two simplifing
assumptions: a) the self-energies are calculated at zero momentum and 
b) we do not distinguish between quarks and gluons.
The latter assumption reduces the number of generic two loop self-energy
diagrams from five to two, shown in Fig.6.
The first type, the rainbow diagram, is shown for $\Sigma^{--}_R$ and 
$\Sigma^{-+}_R$ in more detail in Fig.7. From these two self-energies, the
retarded self-energy can be constructed as
\begin{equation}
\Sigma^r_R = \Sigma^{--}_R + \Sigma^{-+}_R.
\end{equation}
Since the retarded self-energy is a physically relevant property, that
enters, for example, into the constraint equation Eq.(\ref{constraint}), 
we want to investigate the diagrams of Fig.7 in more detail. In each 
of the diagrams b), c) and f) of Fig.7, it is possible to identify an
internal vertex to which lines three 
off-diagonal Green functions are attached. Since these Green functions are
on mass shell, this corresponds to a decay of an on-shell 
particle into two on-shell particles of the same species.
This is a forbidden process, and therefore these three diagrams vanish.
Closer inspection of Fig.7 g) shows that the momentum structure of the Green
functions is similar to that of b), c) and f), and this graph is therefore
also vanishing. Consequently, 
$\Sigma^r_R$ is the sum of the remaining diagrams a), d), e) and h):
\begin{eqnarray}
\Sigma^r_R(X,0) &=& (-igm)^4 \int\!\frac{d^4k}{(2\pi)^4}\frac{d^4l}{(2\pi)^4}
\left\{[D^{--}(k)]^3 D^{--}(k-l)D^{--}(l)\right.\nonumber\\
&&\qquad\qquad\qquad + D^{--}(k) [D^{-+}(k)]^2 D^{++}(k-l) D^{++}(l)\nonumber\\
&&\qquad\qquad\qquad - D^{--}(k) [D^{-+}(k)]^2 D^{--}(k-l) D^{--}(l)\nonumber\\
&&\left.\qquad\qquad\qquad - D^{++}(k) [D^{-+}(k)]^2 D^{++}(k-l) D^{++}(l)
\right\}.
\label{e:r}
\end{eqnarray}
This construction appears to contain pinch singularities, evidenced by the
fact that products of $D(k)$ occur, and it is imperative to show that the
apparent divergence vanishes through cancellation. For simplicity, we
consider this problem first in equilibrium. We do this for two reasons.
Firstly, a correct calculation in equilibrium should be entirely free of
pinch singularities, as has been demonstrated generally in \cite{landsman}.
In addition, calculations can be simplified dramatically, in such a fashion
as to allow for an extension to systems slightly off equilibrium in a simple
fashion in this direct example. We use the following reasoning:  
in equilibrium, the
Schwinger-Keldysh contour is simply one of a family of choices of contour in
which the upper and lower integration routes are separated by a distance
$\sigma$, and with which a form of real time field theory [RTFT] is generated 
\cite{banff}.   In the Schwinger-Keldysh case, one makes the choice $\sigma=0$.
Physical quantites, when calculated, should however be independent of the 
value of $\sigma$.   As it turns out, the particular value $\sigma=1/2$, which
generates the so-called thermo field theory (TFT) formalism \cite{umezawa2}, 
gives distinct calculational advantages, as products of off-diagonal elements 
can be seen to simply vanish. Given this equivalence, we are free to choose
to do our analysis in the TFT framework.    

The scalar Green functions, in the TFT formalism, are given as
\begin{eqnarray}
{iD^{--}\;\,iD^{-+}\choose iD^{+-}\;\,iD^{++}}
&=&{D^{1}_0+D_{\beta}\qquad D'_{\beta}\;\;\;\;\choose \;\;\;\;D'_{\beta}\qquad
D^{2}_0+D_{\beta}}\nonumber\\
&=&{\frac{i}{k^2-m^2+i\epsilon}+\frac{2\pi\delta(k^2-m^2)}{e^{\beta|k_0|}-1}
\;\;\;\;2\pi\delta(k^2-m^2)\frac{e^{\beta|k_0|/2}}{e^{\beta|k_0|}-1}
\choose 2\pi\delta(k^2-m^2)\frac{e^{\beta|k_0|/2}}{e^{\beta|k_0|}-1}\;\;\;\;
\frac{-i}{k^2-m^2-i\epsilon}+\frac{2\pi\delta(k^2-m^2)}{e^{\beta|k_0|}-1} },
\label{e:matrix}
\end{eqnarray}
so defining $D_0^1$, $D_0^2$, $D_{\beta}$ and $D_{\beta}'$\footnote{Our
notation introduces a factor $i$ in the first matrix, in order to establish 
consistency with our Schwinger-Keldysh Green functions of Appendix A. 
This differs from the standard use of TFT users \cite{fuji}.}.
Let us compare these Green functions with the ones in Eqs.(\ref{d-+})-
(\ref{d++}). In equilibrium, $f_a(X,p)=f_a(X,-p)=1/(e^{\beta |p_0|}-1)$.
If we do not distinguish between particles and antiparticles, then the
diagonal Green functions in the TFT formalism are equal to the ones in
Eqs.(\ref{d--}) and (\ref{d++}). The off-diagonal Green functions
differ from Eqs.(\ref{d-+}) and (\ref{d+-}). However, in the self-energies 
considered only the product $D^{-+}D^{+-}$ occurs, which is the same in both 
types of formalism. This in itself demonstrates that one may use the Green 
functions of the TFT formalism with impunity in the following discussion.
We consider first the last term of the retarded self-energy in 
Eq.(\ref{e:r}):
\begin{eqnarray}
\int\!\frac{d^4l}{(2\pi)^4}D^{++}(k-l) D^{++}(l) 
&=& \int\!\frac{d^4l}{(2\pi)^4}\left\{ D^{2}_0(k-l)D^{2}_0(l)
     +D^{2}_0(k-l)D_{\beta}(l) \right. \nonumber\\
&&\left. +D_{\beta}(k-l)D^{2}_0(l)
         +D_{\beta}(k-l)D_{\beta}(l) \right\}\nonumber\\
&=& \int\!\frac{d^4l}{(2\pi)^4}\left\{-D^{1}_0(k-l)D^{1}_0(l)
    - D^{1}_0(k-l)D_{\beta}(l) \right. \nonumber\\
&&\left. -D_{\beta}(k-l)D^{1}_0(l)
         +D_{\beta}(k-l)D_{\beta}(l) \right\}\nonumber\\
&=& - \int\!\frac{d^4l}{(2\pi)^4}D^{--}(k-l)D^{--}(l)
\label{e:lint}
\end{eqnarray}
In the first step, we have made use of the relations \cite{fuji}
\begin{equation}
\int\!\frac{d^4l}{(2\pi)^4}D^{1}_0(k-l)D^{1}_0(l)
=-\int\!\frac{d^4l}{(2\pi)^4} D^{2}_0(k-l)D^{2}_0(l)
\end{equation}
and 
\begin{equation}
\int\!\frac{d^4l}{(2\pi)^4}D^{1}_0(k-l)D_{\beta}(l)
=-\int\!\frac{d^4l}{(2\pi)^4}D^{2}_0(k-l)D_{\beta}(l).
\end{equation}
In the second step, we have made use of the fact that the fourth term,
$D_{\beta}(k-l)D_{\beta}(l)$, has to vanish, since together with
$D'_{\beta}(k)$ of Eq.(\ref{e:r}) it corresponds again to the decay of an
on-shell particle into two on-shell particles.
With Eq.(\ref{e:lint}), we can rewrite Eq.(\ref{e:r}) for the retarded 
self-energy as 
\begin{eqnarray}
\Sigma^r_R(X,0) &=& (gm)^4 \int\!\frac{d^4k}{(2\pi)^4}\frac{d^4l}{(2\pi)^4}
D^{--}(k-l)D^{--}(l) \nonumber\\
&&\times\left\{[D^{--}(k)]^3-2D^{--}(k) [D^{-+}(k)]^2+D^{++}(k)
[D^{-+}(k)]^2\right\}.
\label{e:r2}
\end{eqnarray}
It is now convenient to make use of the following representation of 
$\delta$-function,
\begin{eqnarray}
2\pi \delta(k^2-m^2)&=&\frac{i}{k^2-m^2+i\epsilon}
-\frac{i}{k^2-m^2-i\epsilon}\nonumber\\
&=& D^1_0 + D^2_0.
\end{eqnarray}
Then the term in curly brackets of Eq.(\ref{e:r2}) is evaluated to give  
\begin{eqnarray}
\{...\}&=&[D^1_0+(D^1_0+D^2_0)f_B]^3 
+[D_0^2+(D^1_0+D^2_0)f_B](D^1_0+D^2_0)^2 g_B^2\nonumber\\
&&-2[D^1_0+(D^1_0+D^2_0)f_B](D^1_0+D^2_0)^2 g_B^2,
\end{eqnarray}
where $g_B=e^{\beta |k_0|/2}/(e^{\beta |k_0|}-1)$. Using the fact that
 $f_B^2-g_B^2=-f_B$, we find 
\begin{equation}
\{...\}= (D^1_0)^3+[(D^1_0)^3+(D^2_0)^3]f_B.
\end{equation}
This expression is well defined, since no products of $D^1_0D^2_0$ occur any
longer. With the relation 
\begin{eqnarray}
(D^1_0)^3+(D^2_0)^3&=&-\frac{1}{2}
\left(\frac{\partial}{\partial m^2}\right)^2 (D^1_0 + D^2_0)\nonumber\\
&=& -\frac{1}{2}\left(\frac{\partial}{\partial m^2}\right)^2 2\pi
\delta(k^2-m^2),
\end{eqnarray}
we can write the retarded self-energy as
\begin{eqnarray}
\Sigma^r_R(X,0) &=& (gm)^4 \int\!\frac{d^4k}{(2\pi)^4}\frac{d^4l}{(2\pi)^4}
D^{--}(k-l)D^{--}(l) \nonumber\\
&&\times\left\{ (D^1_0)^3-\frac{1}{2}\left(\frac{\partial}
{\partial m^2}\right)^2 2\pi \delta (k^2-m^2) f_B \right\}.
\label{e:r3}
\end{eqnarray}
We conclude that for the retarded rainbow self-energy $\Sigma^r_R$ no pinch
singularities occur.

We now have to consider the cloud self-energy of Fig.6(b).
The retarded self-energy is constructed in a similar way as for the rainbow
diagram. The calculation of such a diagram within the framework of $\phi^3$
theory has been performed in \cite{fuji}. This result can be simply taken
over for our purposes, and we quote the final expression here:
\begin{eqnarray}
\Sigma^r_C(X,0) &=& (gm)^4 \int\!\frac{d^4k}{(2\pi)^4}\frac{d^4l}{(2\pi)^4}
\left[ \left( D_0^1(k)\right)^2+\chi(k)\right] \nonumber\\
&&\times \left[ D_0^1(k-l)+D_{\beta}(k-l)\right]
\left[ \left( D_0^1(l)\right)^2+\chi(l)\right],
\label{e:c}
\end{eqnarray}
where the function 
\begin{eqnarray}
\chi(k)&=&2D_0^1D_{\beta} + D_{\beta}^2- (D_{\beta}')^2\nonumber\\
&=&2P\frac{i}{k^2-m^2}2\pi\delta(k^2-m^2)f_B\nonumber\\ 
&=&i\frac{\partial}{\partial m^2}2\pi\delta(k^2-m^2)f_B
\end{eqnarray}
is free of singularities. Thus we conclude that in equilibrium, 
the retarded self-energies with two loops have no pinch singularities. We
conjecture that the inclusion of different kinds of particles interacting
with each, as in Section \ref{sec:coll}, will not display pinch 
singularities - the actual demonstration of this fact, however, is 
cumbersome.

Let us now investigate the effect of small deviations from equilibrium, 
i.e.~we replace $f_B=1/(e^{\beta |k_0|}-1)$ by $ f_B+\delta f_B$ in
Eq.(\ref{e:matrix}).
In deriving our result for the retarded self-energies $\Sigma_R^r$ and
$\Sigma_C^r$, Eqs.(\ref{e:r3}) and (\ref{e:c}), a central feature is that 
the relationship $f_B^2-g_B^2=-f_B$ was used. For small deviations 
from equilibrium, this becomes
\begin{eqnarray}
f_B^2 &\to& [f_B+\delta f_B]^2\nonumber\\
g_B^2=f_B(1+f_B) &\to& [f_B+\delta f_B][1+f_B+\delta f_B]\nonumber\\
=> f_B^2-g_B^2 &\to& [f_B+\delta f_B][f_B+\delta f_B-1-f_B-\delta f_B]
\nonumber\\
&=&-[f_B+\delta f_B],
\end{eqnarray}
i.e.~it is still valid. Therefore we obtain for $\Sigma_R^r$ and
$\Sigma_C^r$ expressions such as in Eqs.(\ref{e:r3}) and (\ref{e:c}),
respectively, with $f_B$ replaced by $f_B+\delta f_B$.
Consequently,  even for small deviations from equilibrium, no pinch
singularities occur for the retarded self-energies up to two loops.

\section{Summary and conclusions}
\label{sec:sum}
In this paper, we have demonstrated that for the two loop self-energy, the
graphs which are generated in the non-equilibrium Schwinger-Keldysh
formalism which do not contribute to two $\to$ two processes are either
vanishing or are necessary to renormalize the graphs of one loop order to
the same level of the coupling strength $g^4m^4$. This applies to all
external legs, save the incoming leg, which is fixed externally. This result
can be generalized to $n$ loops: self-energy graphs containing $n$-loops
lead to contributions to the semiclassical Boltzmann equation that are $n\to
n$ in nature plus include all permutations of the particles on the left and
right hand sides \cite{us}. In addition, however, graphs of lower order are
renormalized successively to the same order of the coupling constant.
We have illustrated that, to the two loop level, pinch singularities da not
occur in the equilibrium formulation, as expected from Ref.\cite{landsman} and
by a simple extension of the technique here, for small deviations from
equilibrium.

\section{Acknowledgments}
\label{sec:ack}
One of us (D.S.I.) thanks C. Greiner and S. Leupold for useful comments made
during the course of this study.
\begin{appendix}
\section{Green functions}
In this Appendix, we list the Green functions used in Section
\ref{sec:coll}.
In the Schwinger-Keldysh formalism \cite{schwing} they are defined for the
(scalar) quarks ($D=S$) as
\begin{eqnarray}
iS^c(x,y)=&\left \langle T\phi^{i,l}(x)\phi^{\dagger j,m}(y)\right \rangle
-\left \langle \phi^{i,l}(x)\right \rangle
 \left \langle \phi^{\dagger j,m}(y)\right \rangle&=iS^{--}(x,y)\nonumber\\
iS^a(x,y)=&\left \langle \tilde{T}\phi^{i,l}(x)\phi^{\dagger j,m}(y)\right
\rangle
-\left \langle \phi^{i,l}(x)\right \rangle
 \left \langle \phi^{\dagger j,m}(y)\right \rangle&=iS^{++}(x,y)\nonumber\\
iS^>(x,y)=&\left \langle \phi^{i,l}(x)\phi^{\dagger j,m}(y)\right \rangle
-\left \langle \phi^{i,l}(x)\right \rangle
 \left \langle \phi^{\dagger j,m}(y)\right \rangle&=iS^{+-}(x,y)\nonumber\\
iS^<(x,y)=&\left \langle \phi^{\dagger j,m}(y)\phi^{i,l}(x)\right \rangle
-\left \langle \phi^{i,l}(x)\right \rangle
 \left \langle\phi^{\dagger j,m}(y)\right \rangle&=iS^{-+}(x,y)
\label{e:defs}    
\end{eqnarray}    
and for the (scalar) gluons ($D=G$) as
\begin{eqnarray}  
iG^c(x,y)=&\left \langle T\chi^{a,r}(x)\chi^{b,s}(y)\right \rangle
-\left \langle \chi^{a,r}(x)\right \rangle
 \left \langle \chi^{b,s}(y)\right \rangle
&=iG^{--}(x,y)\nonumber\\
iG^a(x,y)=&\left \langle \tilde{T}\chi^{a,r}(x)\chi^{b,s}(y)\right \rangle
-\left \langle \chi^{a,r}(x)\right \rangle
 \left \langle \chi^{b,s}(y)\right \rangle &=iG^{++}(x,y)\nonumber\\
iG^>(x,y)=&\left \langle \chi^{a,r}(x)\chi^{b,s}(y)\right \rangle
-\left \langle \chi^{a,r}(x)\right \rangle
 \left \langle \chi^{b,s}(y)\right \rangle
&=iG^{+-}(x,y)\nonumber\\
iG^<(x,y)=&\left \langle \chi^{b,s}(y)\chi^{a,r}(x)\right \rangle
-\left \langle \chi^{a,r}(x)\right \rangle
\left \langle \chi^{b,s}(y)\right \rangle
&=iG^{-+}(x,y).   
\label{e:defg}    
\end{eqnarray}    
Here $T$ and $\tilde{T}$ are the usual time and anti-time ordering operators
respectively. Our convention follows that of Ref.\cite{landau}.
The Wigner Transform of the Green functions is given by
\begin{equation}
D(X,p)=\int\!d^4u\,e^{ip*u}\,D\left(X+\frac{u}{2},X-\frac{u}{2}\right)
\end{equation}
with the center of mass variable $X=x+y$ and the relative distance $u=x-y$.

In a standard fashion, the equations of motion for the Wigner transform of
these Green functions, the so-called transport and constraint equations, 
can be derived \cite{us}. They read
\begin{equation}
-2ip\partial_X D^{-+}(X,p)=I_{\rm coll} + I_{-}^A + I_{-}^R
 \quad\quad {\rm transport}
\label{transport}
\end{equation}
and
\begin{equation}
\left( \frac{1}{2} \Box_X - 2p^2 + 2M^2\right) D^{-+}(X,p) = I_{\rm coll} +
I_{+}^A + I_{+}^R
\quad\quad {\rm constraint,}
\label{constraint}
\end{equation}
respectively,
where $M$ is a generic parton mass, $M=m$ for the gluons and $M=0$ for the
quarks. $I_{\rm coll}$ is the collision term 
\begin{eqnarray}
I_{\rm coll} &=& \Pi^{-+}(X,p)\hat{\Lambda}D^{+-}(X,p)
                 -\Pi^{+-}(X,p)\hat{\Lambda}D^{-+}(X,p)\nonumber\\
             &=& I_{\rm coll}^{\rm gain} - I_{\rm coll}^{\rm loss}.
\label{e:collision}
\end{eqnarray}
$I_{\mp}^R$ and $I_{\mp}^A$ are  terms containing
retarded and advanced components respectively,
\begin{equation}
I_{\mp}^R = -\Pi^{-+}(X,p)\hat{\Lambda}D^{R}(X,p)
            \pm    D^{R}(X,p)\hat{\Lambda}\Pi^{-+}(X,p)
\label{e:rim}
\end{equation}
and
\begin{equation}
I_{\mp}^A = \Pi^{A}(X,p)\hat{\Lambda}D^{-+}(X,p)
            \mp    D^{-+}(X,p)\hat{\Lambda}\Pi^{A}(X,p).
\label{e:pim}
\end{equation}
In Eqs.(\ref{e:collision}) to (\ref{e:pim}), the operator $\hat \Lambda$ is
given by
\begin{equation}
\hat{\Lambda}:={\rm exp}\left\{ \frac{-i}{2}\left(\overleftarrow{\partial}_X
 \overrightarrow{\partial}_p
-\overleftarrow{\partial}_p\overrightarrow{\partial}_X\right) \right\}
\label{lambda1}
\end{equation}
and it is set to 1 in the following calculations, which corresponds to the
semi-classical approximation obtained ($\hbar =1$).

For the calculation of the self-energies, it is useful
to introduce the quasiparticle approximation, in which a
{\it free} scalar parton of mass $M$ is assigned the Green functions
\begin{eqnarray}                                          
iD^{-+}(X,p)&=&\frac{\pi}{E_p}\{\delta(E_p-p^0)f_a(X,p)+\delta(E_p+p^0)
\bar{f}_{\bar a} (X,-p)\}\label{d-+}\\
iD^{+-}(X,p)&=&\frac{\pi}{E_p}\{\delta(E_p-p^0)\bar{f}_a(X,p)+\delta(E_p+p^0)f
_{\bar a}(X,-p)\}\label{d+-}\\                            
iD^{--}(X,p)&=&\frac{i}{p^2-M^2+i\epsilon}+\Theta(-p^0)iD^{+-}(X,p)+
\Theta(p^0)iD^{-+}(X,p)\nonumber\\                        
&=&\frac{i}{p^2-M^2+i\epsilon}+\frac{\pi}{E_p}\{\delta(E_p-p^0)f_a(X,p)+
\delta(E_p+p^0)f_{\bar a}(X,-p)\}\nonumber \\ \label{d--}\\
iD^{++}(X,p)&=&\frac{-i}{p^2-M^2-i\epsilon}+\Theta(-p^0)iD^{+-}(X,p)+
\Theta(p^0)iD^{-+}(X,p)\nonumber\\                        
&=&\frac{-i}{p^2-M^2-i\epsilon}+\frac{\pi}{E_p}\{\delta(E_p-p^0)f_a(X,p)+
\delta(E_p+p^0)f_{\bar a}(X,-p)\}\nonumber \\ \label{d++} 
\end{eqnarray}
with $E_p^2=p^2+M^2$, and
which are given in terms of the corresponding scalar quark and gluon
distribution function,
$f_a(X,p)$, and $\bar f_a = 1+f_a$, where $a$ denotes the parton type
$a=q,g$.
        
\end{appendix}

\begin{center}

\begin{figure}[h]
\centerline{\scalebox{1.0}{\includegraphics{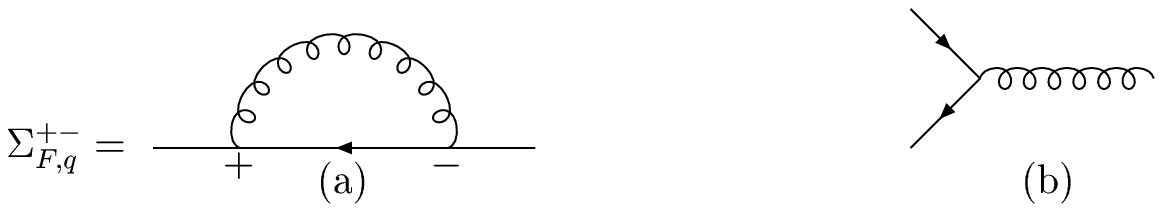}}}
\caption{(a) Fock self-energy diagram and (b) Feynman diagram for the
process $q\bar q\to g$ in lowest order.}
\end{figure}

\begin{figure}[h]
\centerline{\scalebox{1.0}{\includegraphics{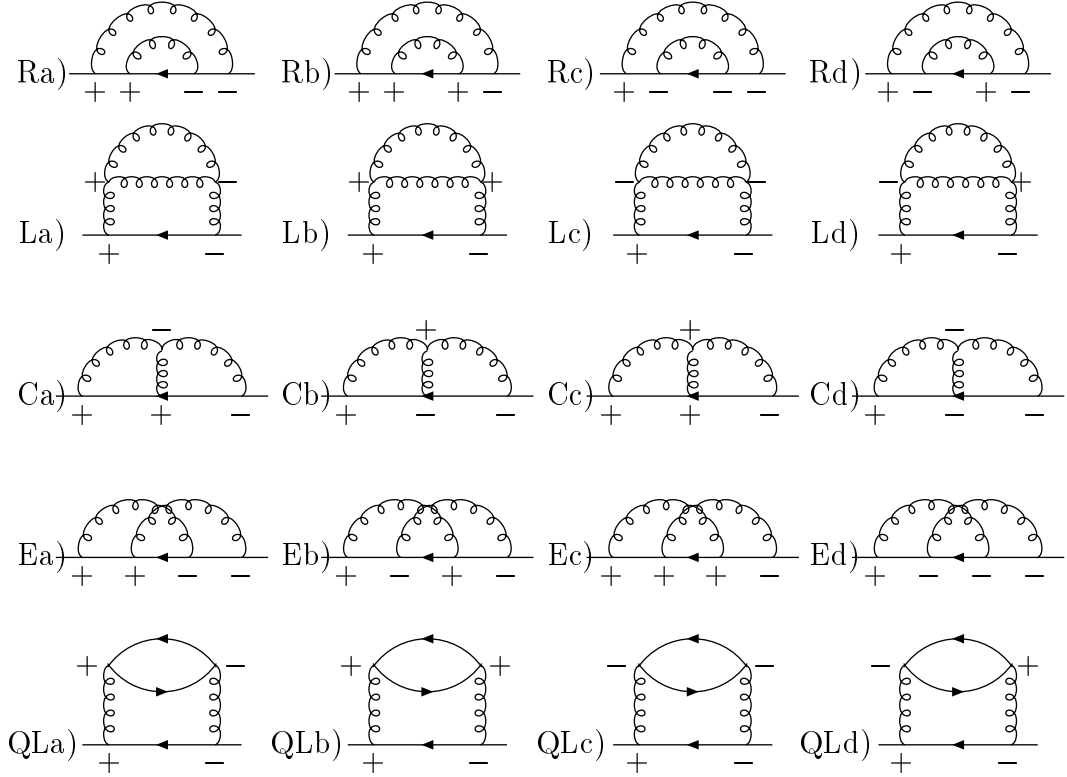}}}
\caption{Two loop self-energy diagrams contributing to
$\Sigma^{(2)+-}$.}
\end{figure}

\begin{figure}[h]
\centerline{\scalebox{1.0}{\includegraphics{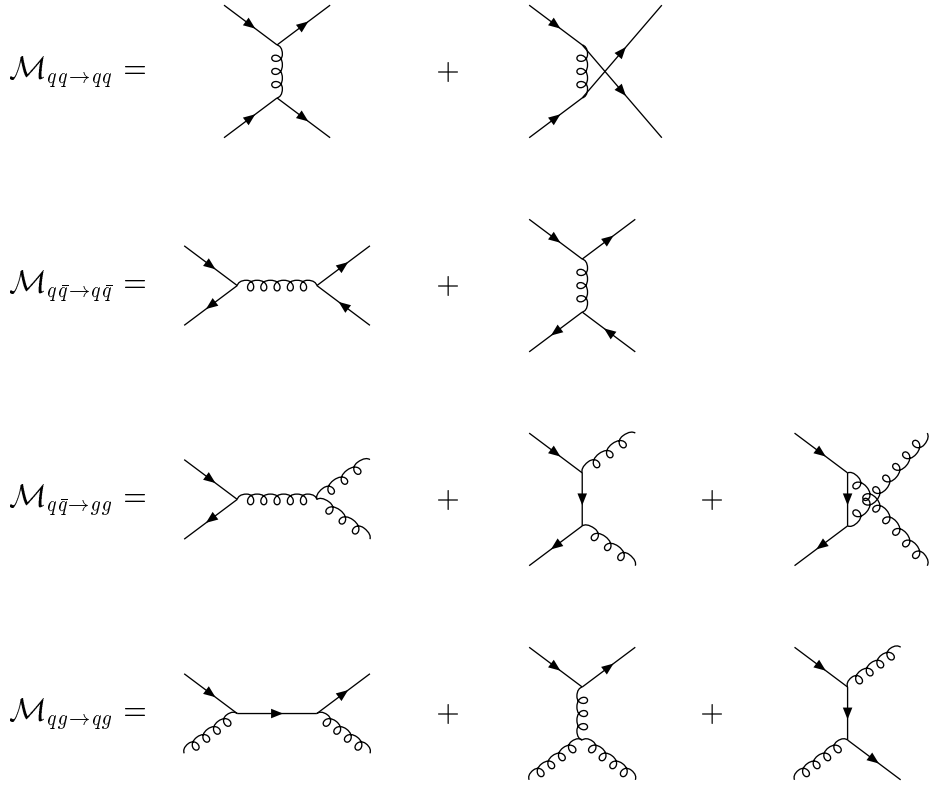}}}
\caption{Scattering amplitudes for all two $\to$ two processes.}
\end{figure}

\begin{figure}[h]
\centerline{\scalebox{1.0}{\includegraphics{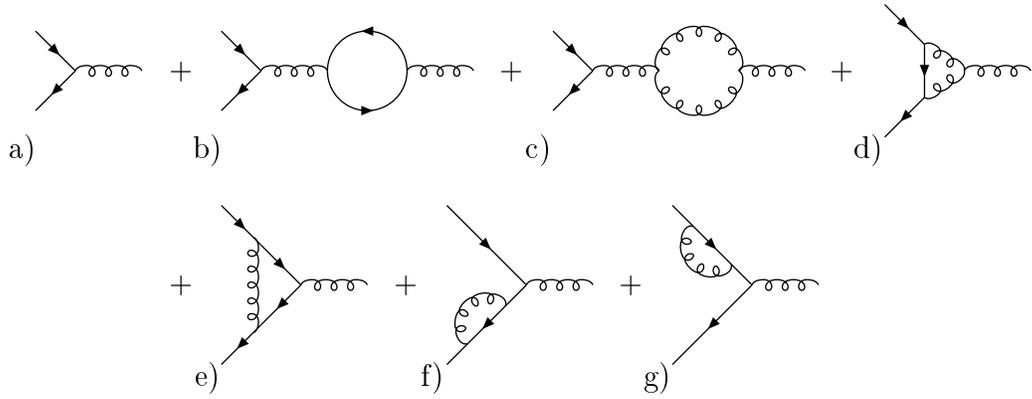}}}
\caption{Feynman diagrams for the process $q\bar q\to g$ up to order
$g^3m^3$.}
\end{figure}

\begin{figure}[h]
\centerline{\scalebox{1.0}{\includegraphics{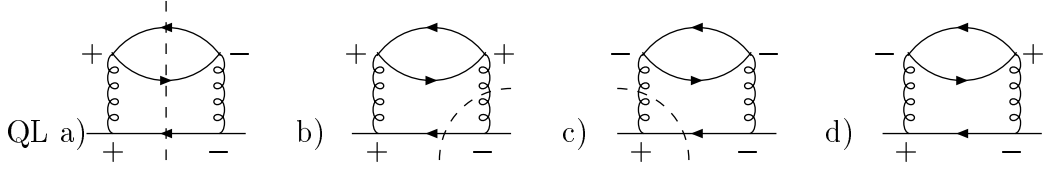}}}
\caption{Quark-loop self-energy diagrams with cut lines (dashed lines).}
\end{figure}

\begin{figure}[h]
\centerline{\scalebox{1.0}{\includegraphics{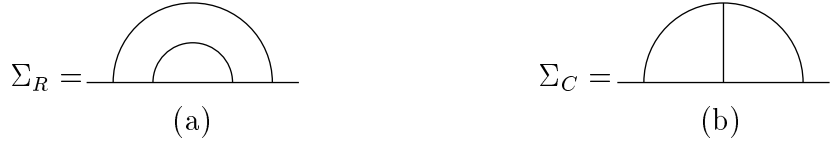}}}
\caption{Generic two loop self-energy diagrams for one parton type.}
\end{figure}

\begin{figure}[h]
\centerline{\scalebox{1.0}{\includegraphics{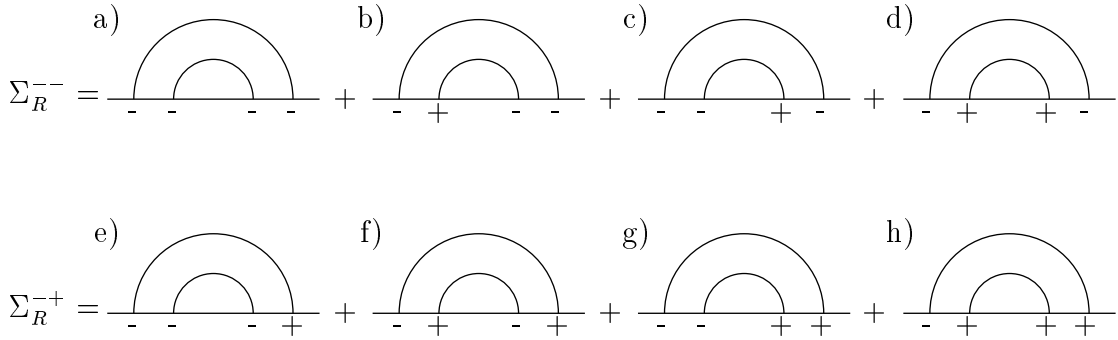}}}
\caption{Rainbow self-energy diagrams for $\Sigma^{--}_R$ and
$\Sigma^{-+}_R$.}
\end{figure}

\end{center}
                                            
\end{document}